# Scoring Grant Applications with Large Language Models

**Mike Thelwall**, School of Information, Journalism and Communication, University of Sheffield, UK. https://orcid.org/0000-0001-6065-205X;
Email: m.a.thelwall@sheffield.ac.uk
**David Pride**, Knowledge Media Institute, The Open University, UK.
https://orcid.org/0000-0002-7162-7252; Email: david.pride@open.ac.uk
**Francesco Osborne**, Knowledge Media Institute, The Open University, UK and Department of Business and Law, University of Milano-Bicocca, Italy.
https://orcid.org/0000-0001-6557-3131; Email: francesco.osborne@open.ac.uk
**Angelo Salatino**, Knowledge Media Institute, The Open University,
UK. https://orcid.org/0000-0002-4763-3943; Email: angelo.salatino@open.ac.uk
**Petr Knoth**, Knowledge Media institute, The Open University, UK.
https://orcid.org/0000-0003-1161-7359; Email: petr.knoth@open.ac.uk
**Gloria Iyawa**, Department of Computer Science and Software Engineering, University of Salford, UK. https://orcid.org/0000-0003-1059-6656; Email: g.e.iyawa@salford.ac.uk
**Oluwaremilekun Yusuf**, Department of Computer Science and Software Engineering, University of Salford, UK. https://orcid.org/0009-0009-7766-5995;
Email: O.Yusuf1@salford.ac.uk

**Purpose**: Assessing grant applications is time-consuming and difficult, adding to the overall burden of academic peer review. Whilst funders are exploring whether AI can help, there is no published research into the accuracy of Large Language Models (LLMs) for scoring contemporary grants.
**Design/methodology/approach**: This study investigates whether six open-weight LLMs (Gemma 3 1B/4B/12B/27B, DeepSeek R1 32B, Qwen 3 32B) can give useful scores for 2267 recent UK Economic and Social Research Council (ESRC), and Engineering and Physical Sciences Research Council (EPSRC) grant applications, comparing them with scores from the original reviewers and funding panel members.
**Findings**: Although the LLM scores are individually inaccurate, when averaged and converted to ranks they correlate positively with expert average scores. The best performing LLM, Gemma 3 27B (10 iterations with varied prompts), had moderate rank correlations with average reviewer scores (mean rho=0.26). Gemma 3 27B's average correlation with individual reviewers was 0.19, which is lower than the inter-reviewer mean correlation of 0.24, suggesting that it scores are slightly weaker than individual reviewer scores. Gemma 3 27B had weak rank correlations with average panel member scores (mean rho=0.17), with lower average correlations with individual panellists (mean rho=0.14), which is substantially lower than the inter-panellist correlation (mean rho=0.38).
**Research limitations**: The data covers one type of grant application from one country and a limited set of LLMs. Only the vision and approach sections were available for the applications.
**Practical implications**: Whilst the correlations seem too weak to replace expert review at the final panel stage, LLM scores might help with the initial reviewing state, such as by helping identify the weakest proposals for fast-track desk rejections, to replace one human reviewer, or for triangulation to check for bias.

**Originality/value**: This is the first analysis of the usefulness of LLM scores for contemporary general research grants.
**Keywords**: Large Language Models; Funding application; Scientometrics; Bibliometrics; Research evaluation

# 1 Introduction

Many countries offer competitive funding opportunities to researchers, as do charities and other private organisations. The selection process typically involves human review by academic peers or other experts or stakeholders. This is time-consuming and error-prone because applications can be long and complex and may need a range of types of expertise to fully evaluate, such as for the methods, topic, planning, and dissemination. Because of this, inter-reviewer agreement rates tend not to be high (e.g., a correlation of 0.2 for 4,000 social science proposals in: Jerrim and De Vries 2023; see also: Carpenter and Corvillón 2026; Tamblyn et al. 2023) and the process can be slow as well as adding to the academic peer review burden (Aczel et al. 2021). In response, the funder La Caixa uses Large Language Models (LLMs) for application triage in the sense of identifying and rejecting proposals that are unlikely to be successful, supported by fast-track human review to safeguard against mistakes (Carbonell Cortés et al. 2024). Evidence of the accuracy of LLMs for this task, as well as the best way of using them, is needed to help other funders decide whether LLMs could also be useful for this or other roles. Corporate applications to government and citizen financial requests are already sometimes assessed with the help of AI, apparently increasing efficiency without reducing accuracy (Marques et al. 2025), and this would be an ideal outcome for funders.

## *1.1 Is there a gold standard for grant proposal quality scores?*

A fundamental issue with peer and expert decisions is that they are subjective with no gold standard of "correct" scores or outcomes. For example, if experts almost always agreed on the merits of grant proposals in their field (e.g., as expressed by a six-point score scale), then expert scores might be uncontroversial as a ground truth. Nevertheless, experts often sharply disagree, emphasising the subjective nature of the issue (Pier et al. 2018). For example, in fields with competing paradigms, the score for a proposal might depend substantially on whether it aligns with the expert's beliefs. Moreover, reviewers tend to be harsher on proposals that they are more expert about (Gallo et al. 2016), so proposals that are not reviewed by anyone with direct expertise have an advantage. There is not an obvious alternative way of obtaining a score, however. Even for retrospective evaluations, if funded proposals were judged on their outcomes (e.g., Li and Agha 2015), then this would be unfair to proposals that had reasonable ideas that did not work for reasons out of the researchers' control, such as an idea for a vaccine that had to be abandoned due to unexpected side effects. Academic research fundamentally involves uncertainty, so it is not possible through any methods to predict the most successful research to fund in advance (Roumbanis 2021). Scoring research proposals therefore involves, in part, estimating the likelihood of success and the value of that success.

The standard academic funder response to the grant reviewing problem, at least in the UK, is to solicit reviews and scores from multiple field experts and then convene a decision panel of different experts to select the proposals to be funded, guided by these scores (Jerrim and De Vries 2023). This encodes the belief that field experts' opinions are the best source of evidence about the quality of proposals and that the average score given by experts, mediated by an expert decision panel, is the closest to a correct answer that it is practical to obtain (e.g., Fogelholm et al. 2012). The role of the decision panel deliberation might be to discard obviously biased, inexpert or incorrect reviews, to inform judgements about what is best for the wider health of the field (e.g., to fund equal numbers of qualitative and quantitative proposals), or to review the decision about which proposals to fund when this is influenced by score ties or where there is only a small difference in scores for proposals near the funding cut-off threshold (Gallo et al. 2020; Graves et al. 2011; Oxley 2025).

In recognition of the subjective nature of the process, the time-based cost of peer review, and the potential for nepotism or group-based biases, some funding calls use partial random selection (e.g., for all proposals passing an initial peer review quality threshold) (Roumbanis 2019). If another source of evidence, such as LLMs, was able to produce scores that tended to agree with expert scores then they would produce results that statistically funded better proposals than random processes and, depending on the level of agreement with expert scores, might approach the level of presumed human expert success at giving higher scores to better proposals.

### *1.2 The value of LLMs for academic reviewing tasks*

Although previous studies have found that LLMs can give useful reviewer suggestions for some conference submissions (Thakkar et al. 2025) and can generate research quality scores for submitted or published research papers that correlate positively with expert scores (Thelwall and Yaghi 2025), few studies have used LLMs to score research proposals. This is presumably because the reviewer scores and unsuccessful grant proposals are usually confidential. In related research, LLMs have been used to generate evaluative comments on six research proposals: they were not good enough to replace humans but provided some useful suggestions for reviewers (Thorne et al. 2026). An analysis of proposals for the high energy physics laboratory Fermilab found that traditional machine learning had a good ability to predict a range of project outcome indicators (Sikimić and Radovanović 2022), but expert scores were not compared. More generally, a different approach showed that project outcome indicators could help to predict future income levels for scientific topics (Jenset 2025). In terms of scoring proposals, one study focused on a single physics facility found that unnamed LLM scores aligned strongly with expert scores (Ding et al. 2025).

In contrast to the above cases, there has been one published attempt to use LLMs to score national funding proposals, albeit on a small scale and with old data: open-weight LLMs scored 142 Swedish medical fellowship research proposals from 1994 that had been released as part of a freedom of information request. The proposals were scanned from printed documents, and all Swedish text was translated to English. Based on intuitive prompts submitted five times per proposal and model, there were moderate rank correlations with expert scores: a mean Spearman's rho = 0.33 for Gemma 3 27B processing proposal titles and summaries, falling to 0.24 when processing full texts. The correlations were positive but generally weaker for DeepSeek R1 70B (0.23 for title/summary, 0.21 for full texts), DeepSeek R1 32B (0.21 for title/summary, 0.12 for full texts), Llama4 Scout (0.21 for title/summary, full texts not processed), Magistral Small (0.24 for title/summary, 0.15 for full texts), and Qwen 3 32B (0.25 for title/summary, 0.25 for full texts) (Sandström and Thelwall 2026). These estimates are relatively inaccurate, however, due to the small sample size. Moreover, the applications are 32 years old and are restricted to one country and application type. Another study used an unnamed LLM for proposals from India but did not explain the methods in detail (Nagarajappa et al. 2026).

### *1.3 Research questions*

The current study addresses the absence of knowledge about how well LLMs can score contemporary national grant proposals and adds to the minimal knowledge so far about how LLMs can score any grant proposals. Since investigations into scoring journal articles for quality have found that the scores are inaccurate overall, but the rankings can still align between humans and LLMs, proposal ranking is the focus of the paper. In addition, the LLMs are restricted to the open-weight type that can be downloaded and used offline in secure environments, which is sometimes necessary to preserve confidentiality (and was a funder condition of the current study). They are also restricted to medium-sized LLMs, since the largest open-weight LLMs need expensive specialist equipment to run at a reasonable speed. This study therefore addresses the following research questions.

- RQ1: How strongly do average scores from small or medium sized open-weight LLMs correlate with expert scores for contemporary grant funding proposals?
- RQ2: Are there differences between funders in the answers to RQ1?

## 2 Methods

The research design was to correlate average scores for grant proposals from a range of small and medium-sized LLMs with reviewer scores for a set of funding calls. A range of LLMs was used rather than one because performance can vary between LLMs for the same task, for example due to size or architecture. Ethical approval was received for this study from the University of Sheffield (069782).

### *2.1 Data*

The vision and approach sections of grant proposals were obtained for four calls from UK Research and Innovation (UKRI), who oversee UK government academic research funding. These were all open calls (no specific topic, called “responsive mode” by UKRI) from 2023 with two from the Economic and Social Research Council (ESRC) and two from the Engineering and Physical Sciences Research Council (EPSRC). The proposals were shared along with the individual reviewer “scores” and the “prescores” from members of the final funding selection panel. Proposals did not have “prescores” if they were not shortlisted due to low average reviewer scores. There were varying numbers of scores and prescores per application, but, when present, there were usually at least two and no more than six. The (reviewer) scores on an integer scale of 1 to 6 are from academics responding to requests to review individual proposals. The (panellist) prescores on a non-integer scale (one decimal place) of 0 to 10 are from people who sit on the funding panel and would be expected to evaluate multiple applications for each call. Individual panellists are provided with the reviewer scores and comments to inform their initial evaluation of applications and the prescore that they assign. For some panels these pre-scores can then be used to filter out more low scoring applications without panel discussion. During the panel meeting, the selection panel discusses and agrees a final panel score for each application in order to rank the proposals (without knowing the funding cut-off), but this final score was not available due to UKRI system restrictions.

Because of the sensitive nature of the data, the proposals and scores were made available only within a secure cloud-based Linux environment controlled by UKRI. As part of the access conditions, the environment was blocked from the internet (other than for logon access) and software could only be installed by the UKRI technical team, following a request, and installation was sometimes not possible or not practical, which limited the number of analyses that could be conducted.

The proposals were shared in two forms, plain text only (the second ESRC call) and PDF only (one ESRC call; two EPSRC calls). Since contemporary open-weight LLMs do not accept PDF input, the first stage was to convert the PDFs to text. This was achieved with a Python program using GROBID (Lopez 2009).

## *2.2 Large Language Models*

Following previous research scoring academic journal articles (Thelwall and Mohammadi 2026), Gemma 3 27B from Google was selected as the main medium sized open weight LLM when the experiment was designed and UKRI was asked to make it available in the secure environment in November 2025. This choice aligned with subsequent research on Swedish grant proposals (Sandström and Thelwall 2026). For a check of the influence of size, smaller variants from the same family (1B, 4B, 12B) were also tested, where the B value indicates the number of model parameters (billions). To test whether reasoning might help this task, the medium-sized models DeepSeek R1 32B and Qwen 3 32B were also used. Both were relatively recent and could be imported into the secure environment. All LLMs were run through Ollama as a server within the same secure environment.

The second core decision for the LLM was the wording of the prompts to use. This was taken from the assessment questions and criteria associated with the calls on the UKRI website. This information explains the criteria to applicants and reviewers. Calls sometimes have a “highlight notice” for topics that are particularly welcomed (e.g., “AI for Social Science”). This was not incorporated into the instructions because it was unclear whether the reviewers would be aware of it and, if so, the amount of weight that they would give to it. For both EPSRC and ESRC the main assessment criteria were:

- vision
- approach
- data management and sharing
- applicant and team capability to deliver
- ethics and responsible research and innovation (RRI)
- resources and cost justification.

Unfortunately, only the vision and approach sections had been provided for the applications, so the last four criteria lacked the corresponding proposal sections. The criteria were retained, however, on the basis that some relevant information may be in the approach section. In addition, altering the criteria would have created a different evaluation task from that undertaken during peer review. Nevertheless, because the LLM lacked information available to human reviewers, the reported correlations are likely to represent conservative estimates of performance.

Two system prompts were generated for each funder (see Appendix): a short task description that relied on LLM task understanding and a longer one that included more context from the application. Both were used to test if either gave better results.

Improved results from LLMs scoring academic documents can be obtained if (a) multiple scores are averaged (Thelwall 2024) and (b) the wording of the prompts is varied slightly each time (Thelwall 2026). Averaging scores from multiple prompts seems to help because LLMs use probability-based processing when ingesting texts, so can "understand" a text differently each time they read it. This differing understanding can trigger "knowledge" stored in different parts of the language model. Averaging multiple scores can therefore harness more of a LLM's memory. In addition, when the score is reported, this is also probabilistic, and asking multiple times for a score reveals implicit information about how certain the LLM is about the score. For example, if it scores 4 four times and 5 once, then the average 4.2 reflects the proposal being potentially slightly better than 4. Varying the wording of the prompt whilst keeping the same meaning helps the first of these two strategies, increasing the variety of scores (Thelwall 2026). Thus, each user prompt was submitted five times, each with slight wording variations (see Appendix). The scores from the two system prompts were also combined to give an additional score (i.e., the average of ten scores with all different prompts), also potentially having additional prompt variation benefits, as mentioned above. As a result of the above process, each proposal was submitted up to ten times per model due to the two system variations and five user prompt variations. The two reasoning models ran slowly on the system provided, however, so only the long variant of the prompt was used for them.

The above strategy extends that used previously for the 1992 Swedish medical applications (Sandström and Thelwall 2026) by adding prompt variety and system prompt alternatives. Fine-tuning and fewshot prompting were not used because these add to the processing burden without strong prior evidence suggesting that they would work well. For the same reason (Thelwall 2025), only the default LLM configuration parameters were used.

### *2.3 Analysis*

Although most evaluations of AI use standard accuracy metrics, including precision, recall, F1-measure, and mean absolute deviation, none of these are relevant here because LLM scores for academic purposes are known to be inaccurate because they tend to give higher scores than human experts, which assumed to be the most reliable sources of judgement. Nevertheless, LLM scores have value for ranking documents and, if desired, can be mathematically transformed to adjust to the human scale, such as by subtracting and multiplying by correction factors (Thelwall 2024). Thus, the only useful measure of accuracy is the rank correlation between the LLM scores and expert scores, as reflected by the Spearman correlation, and this is also how the final panel scores are used by funders. Thus, this was the sole performance measure used. Confidence intervals were calculated using bootstrapping.

## 3 Results

The LLM scores tended to have different means and standard deviations from the expert scores (Table 1). In particular, all LLMs had a higher mean score and a lower standard deviation than human reviewers for all calls. This confirms that accuracy measures are not appropriate for evaluating the usefulness of the LLM scores for funding proposals.

Table 1. Mean (standard deviation) panel member scores, reviewer prescores, and LLM scores for the two prompt types and their combination.

| Call (proposals) | Score (Prescore) | Prompt | Gemma 3 27b | Gemma 3 12b | Gemma 3 4b | Gemma 3 1b | DeepSeek R1 32b | Qwen 3 32b |
|---|---|---|---|---|---|---|---|---|
| EPSRC 1 (n=524) | 4.37 (0.82) 7.63 (1.04) | long | 4.82 (0.29) | 4.64 (0.27) | 5.29 (0.33) | 5.05 (1.59) | 4.97 (0.24) | 5.11 (0.39) |
| | | short | 4.58 (0.34) | 4.36 (0.23) | 4.97 (0.44) | 4.78 (0.49) | | |
| | | both | 4.70 (0.29) | 4.50 (0.23) | 5.14 (0.37) | 4.92 (0.89) | | |
| ESRC 1 (n=282) | 4.27 (0.81) 8.15 (0.46) | long | 4.82 (0.29) | 4.61 (0.29) | 5.03 (0.41) | 5.04 (0.46) | 5.00 (0.27) | 5.09 (0.38) |
| | | short | 4.41 (0.47) | 4.30 (0.24) | 4.59 (0.52) | 4.80 (0.44) | | |
| | | both | 4.61 (0.35) | 4.45 (0.25) | 4.81 (0.45) | 4.92 (0.43) | | |
| ESRC 2 (n=752) | 4.09 (0.83) 7.89 (0.75) | long | 5.00 (0.23) | 4.75 (0.26) | 5.03 (0.41) | 5.21 (0.77) | 5.04 (0.24) | 5.17 (0.37) |
| | | short | 4.71 (0.33) | 4.45 (0.24) | 4.59 (0.52) | 4.98 (0.34) | | |
| | | both | 4.86 (0.25) | 4.60 (0.23) | 4.81 (0.45) | 5.09 (0.48) | | |
| EPSRC 2 (n=709) | 4.34 (0.87) 7.57 (1.07) | long | 4.84 (0.29) | 4.64 (0.29) | 5.12 (0.33) | 5.21 (0.77) | 4.99 (0.23) | 5.11 (0.42) |
| | | short | 4.63 (0.37) | 4.38 (0.24) | 4.73 (0.45) | 4.98 (0.34) | | |
| | | both | 4.73 (0.31) | 4.51 (0.24) | 4.92 (0.37) | 5.09 (0.48) | | |

The levels of agreement within and between individual initial reviewers (Score) and panellists (Prescore) were calculated to give a benchmark for the LLM agreement rates. Whilst reviewers (Score) agree moderately with each other (average of 0.242), panellists have a stronger level of agreement (average of 0.381) even though their scores are restricted to pre-filtered high-quality proposals (Figure 1). This aligns with previous research showing that panel members tend to agree more than occasional reviewers, perhaps because they interpret the scale more consistently (Fogelholm et al. 2012; Hodgson 1995). Generally, the human agreement rates are higher for EPSRC than for ESRC.

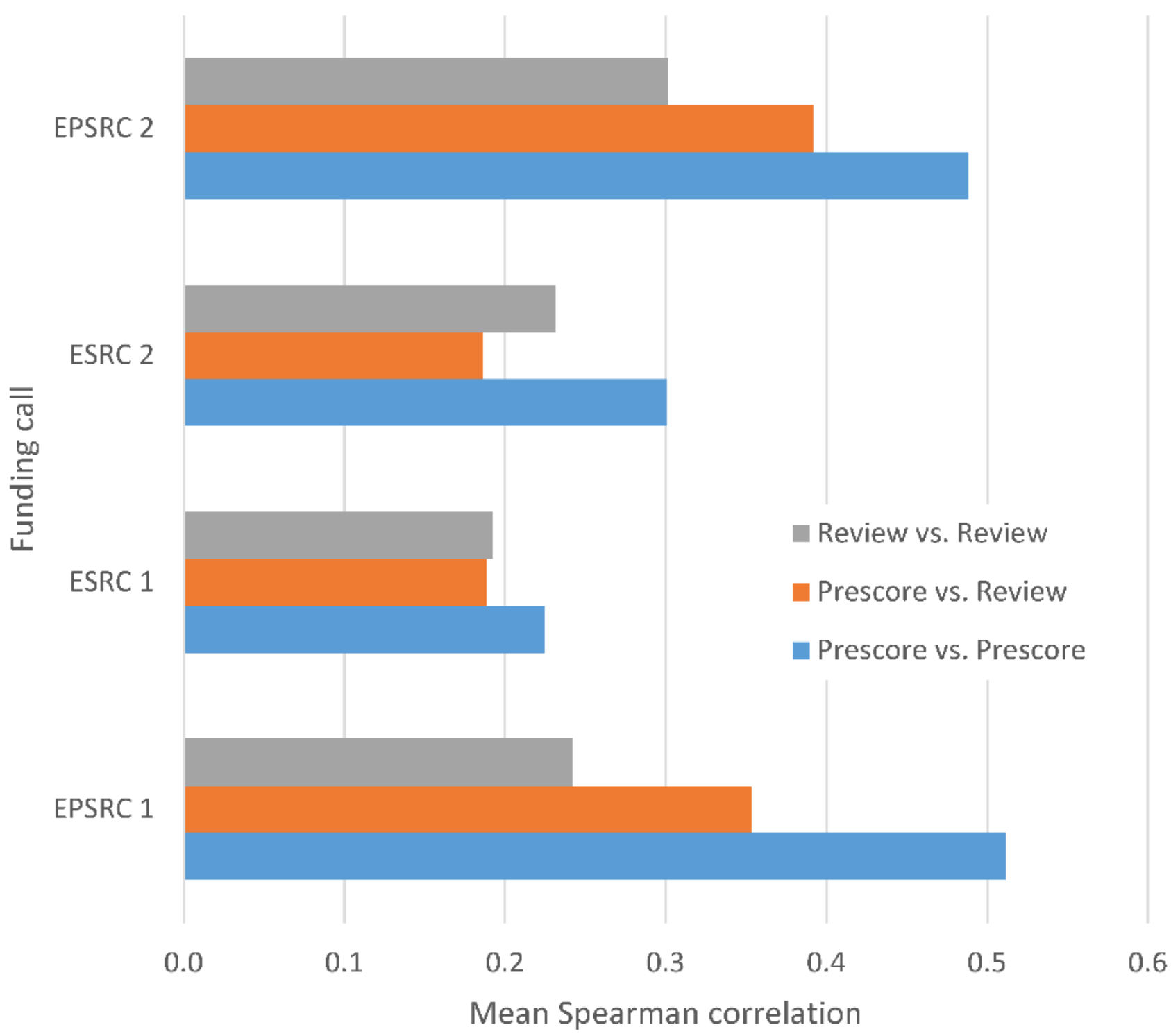

Figure 1. Reviewer and panel member agreement rates, based on the first three reviewers/panellists.

### *3.1 LLM scores vs. initial reviewer scores*

The LLM with the highest correlation with reviewer scores is Gemma 3 27B, and the highest correlation generally occurs when the long and short prompt scores are combined (i.e., the average of 10 scores, 5 from each set of prompts). The mean of the four correlations (one per funding call) for this configuration is moderate, at rho=0.263. Although the correlation confidence intervals are too wide for strong statistical conclusions (Figures 2-5), the two reasoning models perform consistently worse than Gemma 3 27B and the smallest variant of Gemma 3 27B performs substantially worse overall.

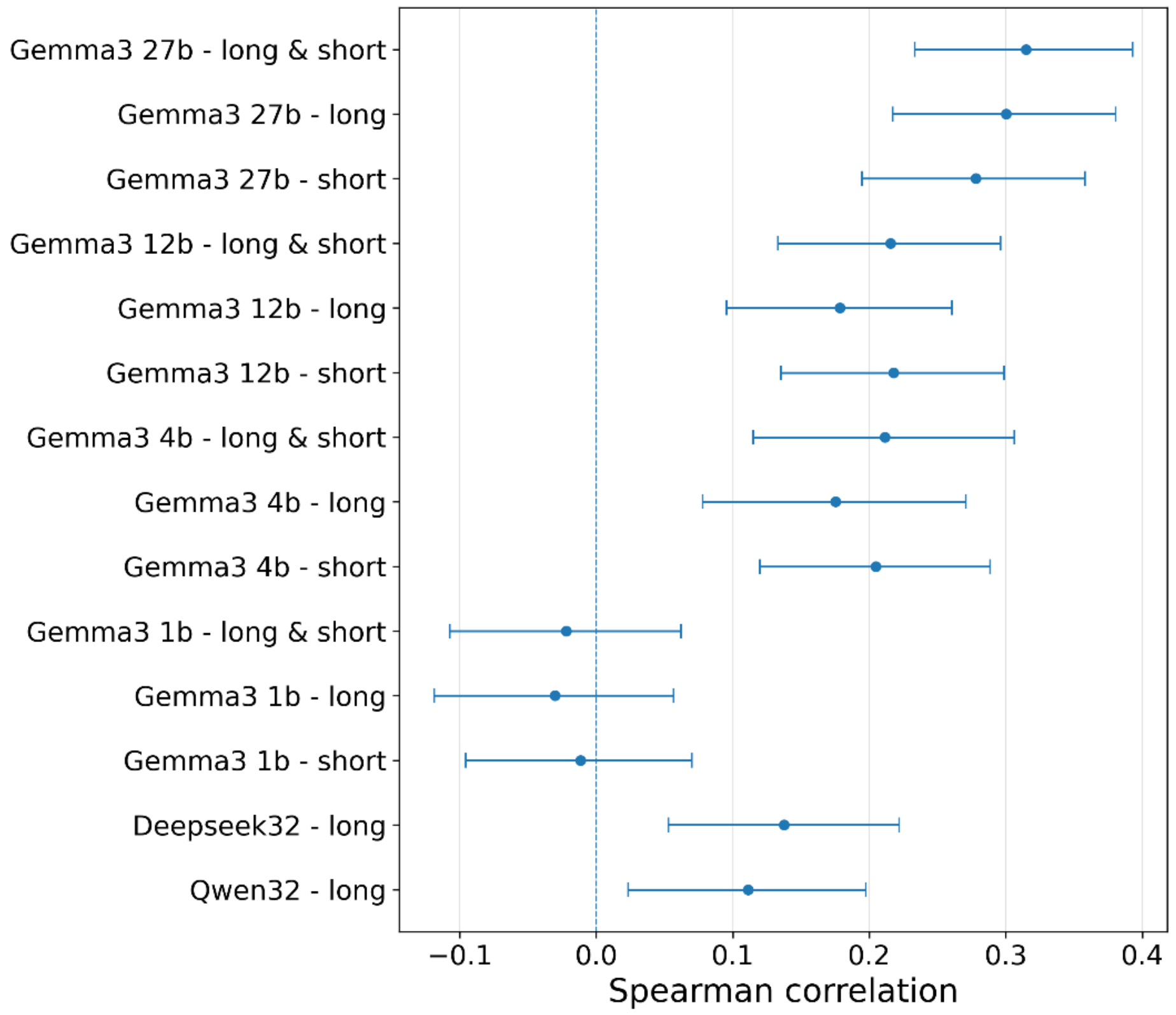

Figure 2. Spearman correlations between LLM scores and reviewer scores for EPSRC 1. Error bars indicate bootstrapped 95% confidence intervals.

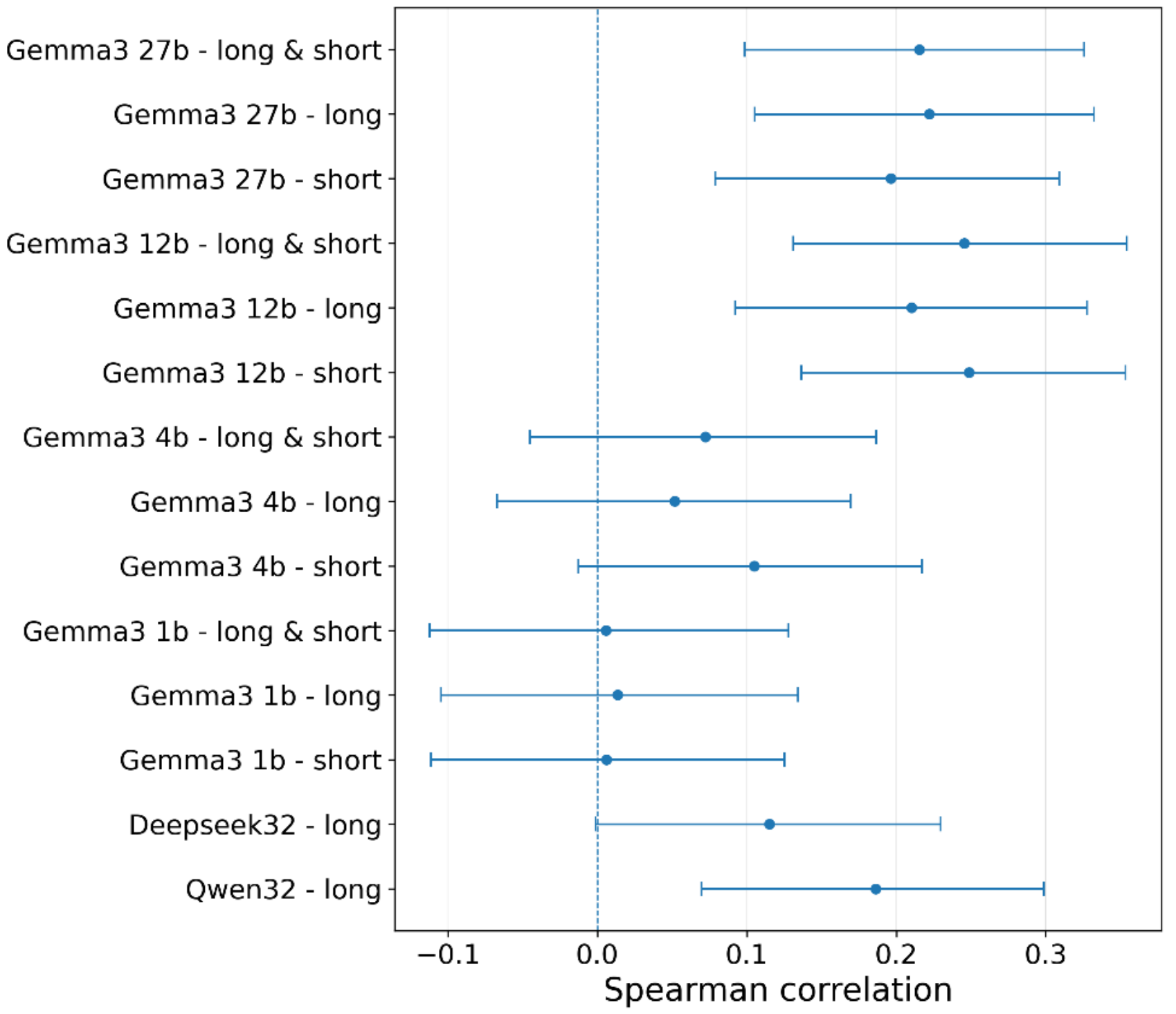

Figure 3. Spearman correlations between LLM scores and reviewer scores for ESRC 1. Error bars indicate bootstrapped 95% confidence intervals.

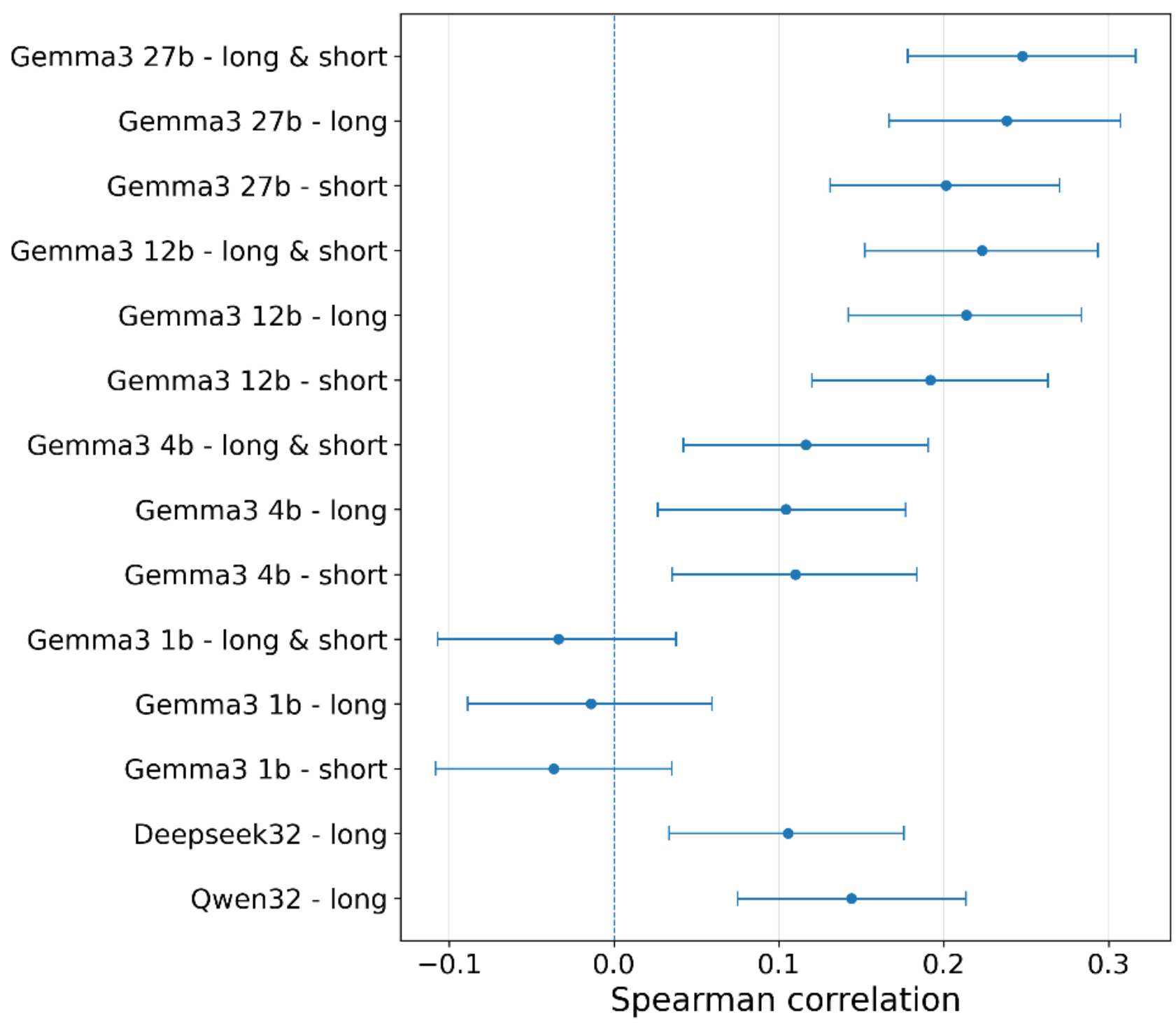

Figure 4. Spearman correlations between LLM scores and reviewer scores for ESRC 2. Error bars indicate bootstrapped 95% confidence intervals.

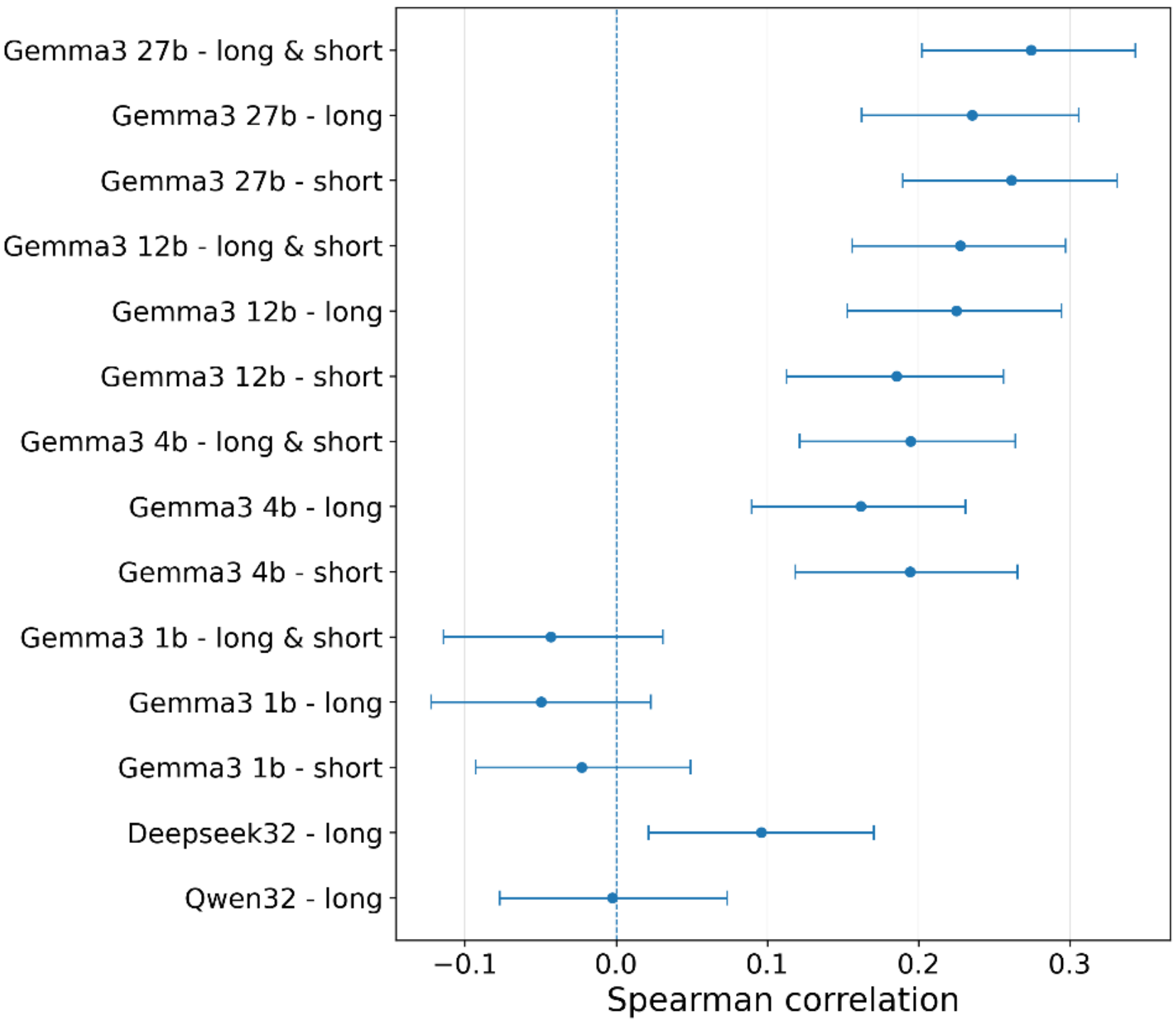

Figure 5. Spearman correlations between LLM scores and reviewer scores for EPSRC 2. Error bars indicate bootstrapped 95% confidence intervals.

### *3.2 LLM scores and panellist prescores*

For panellist prescores, Gemma 3 27B again tended to have the highest correlations, although the differences are not clear-cut and no prompt configuration performed consistently best. Across the four calls, the mean correlation for Gemma 3 27B with the long and short prompt scores combined was weak (rho = 0.165). Given the wide confidence intervals (Figures 6–9), the apparently weaker performance of the two reasoning models and the smaller Gemma 3 variants may be marginal.

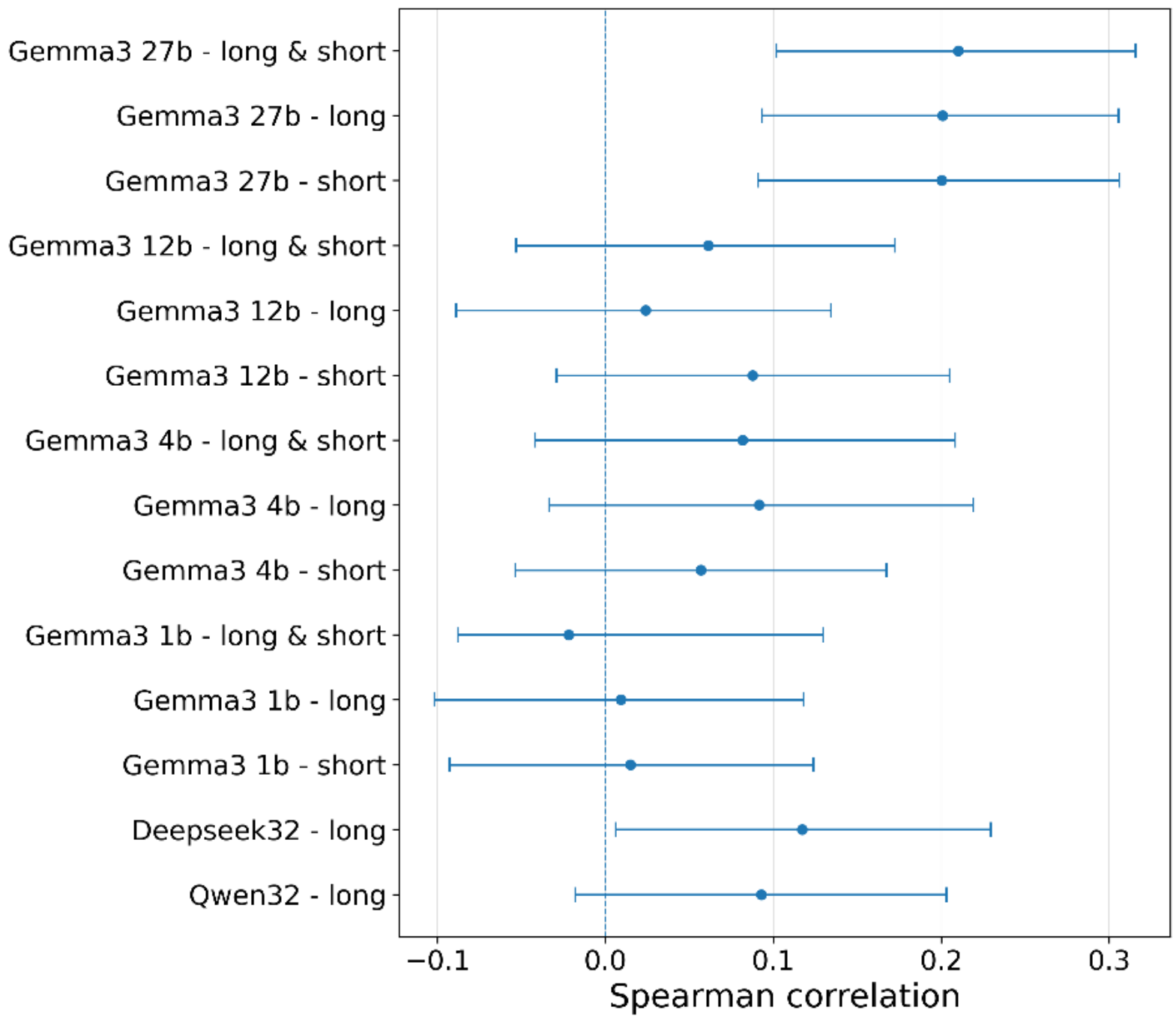

Figure 6. Spearman correlations between LLM scores and panellist prescores for EPSRC 1. Error bars indicate bootstrapped 95% confidence intervals.

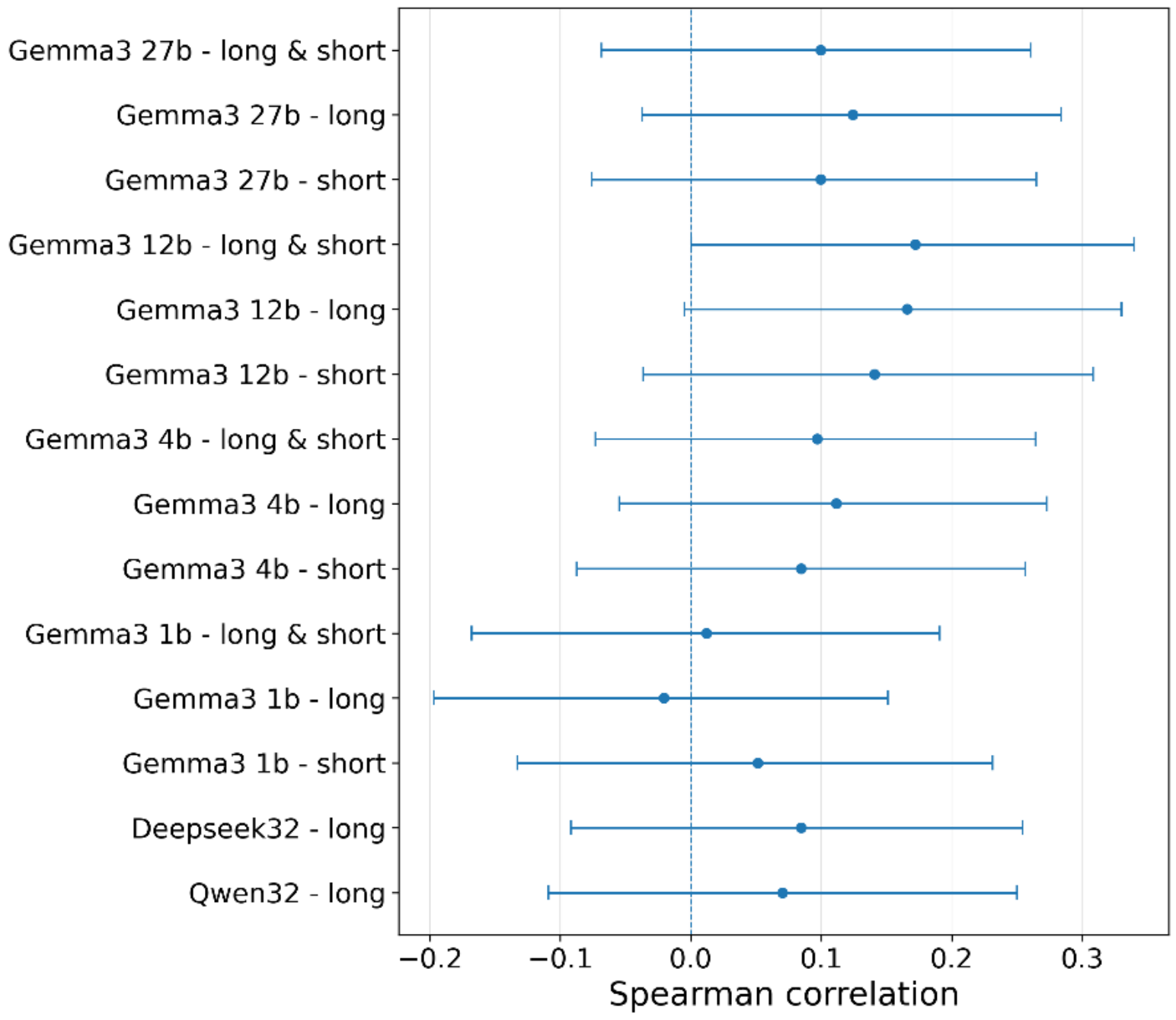

Figure 7. Spearman correlations between LLM scores and panellist prescores for ESRC 1. Error bars indicate bootstrapped 95% confidence intervals.

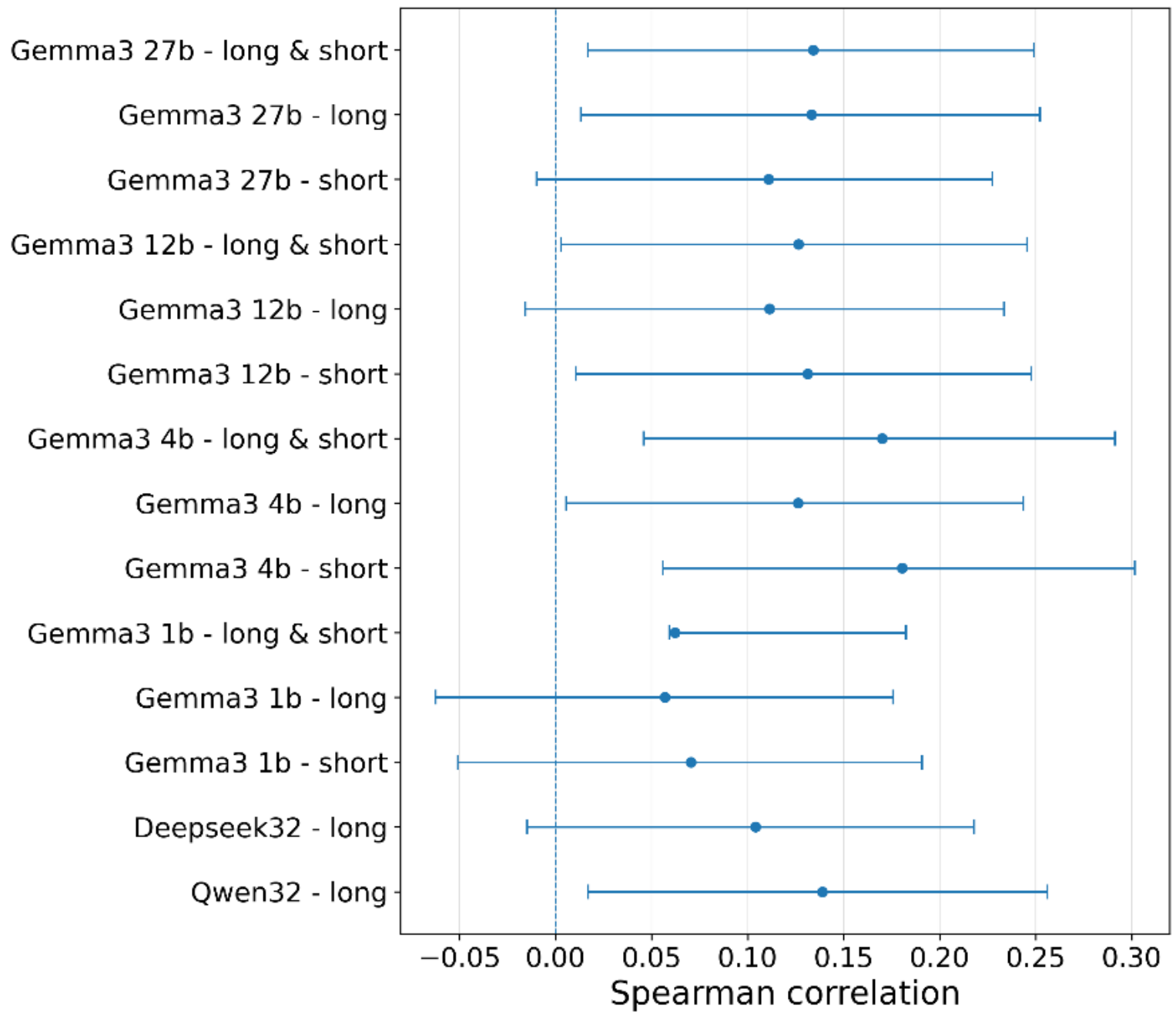

Figure 8. Spearman correlations between LLM scores and panellist prescores for ESRC 2. Error bars indicate bootstrapped 95% confidence intervals.

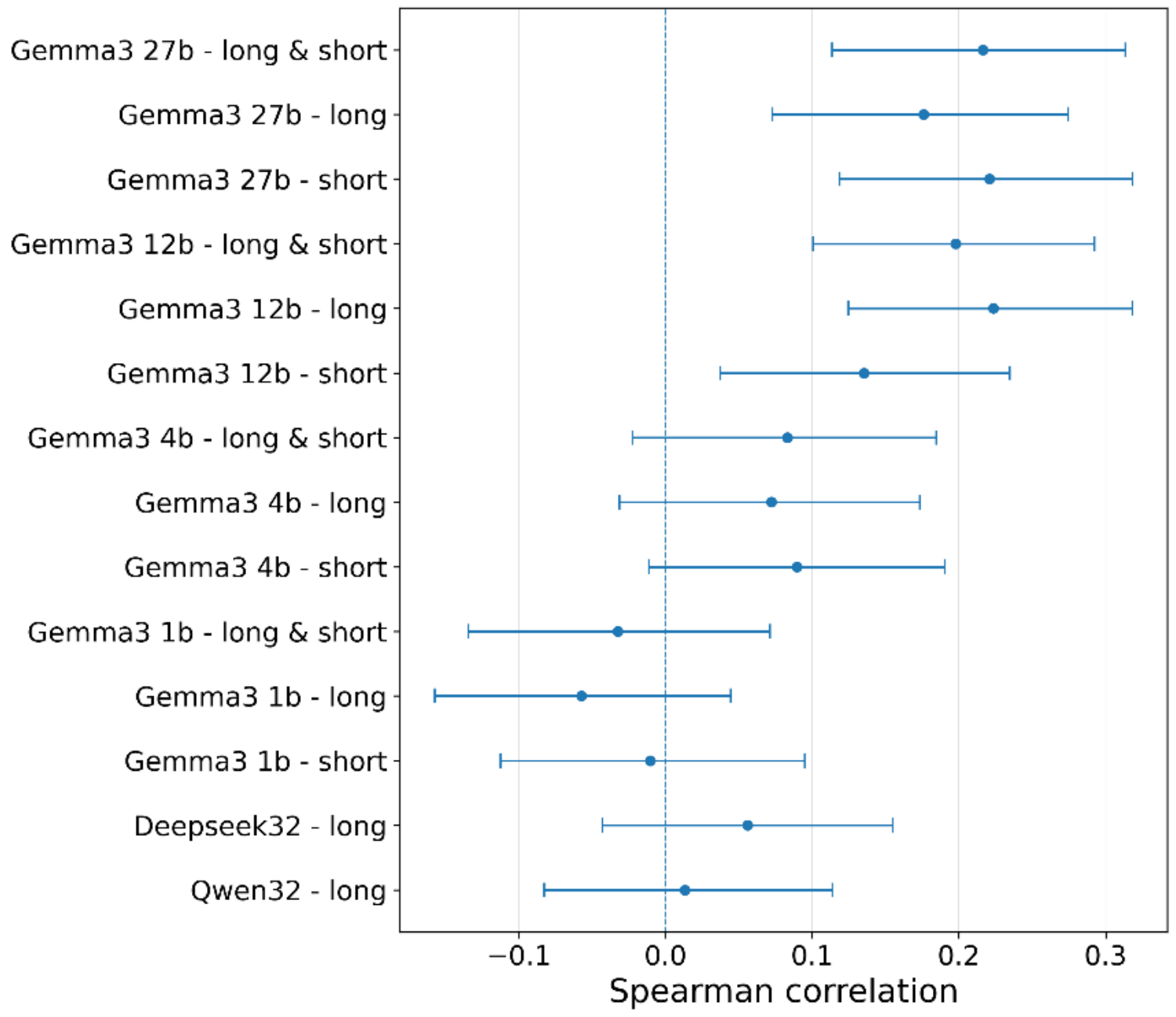

Figure 9. Spearman correlations between LLM scores and panellist prescores for EPSRC 2. Error bars indicate bootstrapped 95% confidence intervals.

# 4 Discussion

The results are limited to one type of funding application (open calls), a single funder type (government), a limited set of LLMs, two areas (social sciences, and engineering and physical sciences), a single year (2023) and one country (the UK). In addition, the results are restricted to two sets of prompts, one prompting strategy (zero shot) and one set of LLM parameters (the defaults). Changing any of these could give different results. Nevertheless, the results are like those from a different year (1994), country (Sweden), field (medicine), and application type (fellowships) (Sandström and Thelwall 2026), which gives some confidence that they are not unique to the context analysed. A substantial limitation is that only proposal texts could be processed and only for the Vision and Approach application sections without the summaries or remaining parts.

A more conceptual limitation is that the analysis assumes that the average human reviewer score is the correct answer, or at least the best available approximation, but it is impossible to know how good reviewers and panel members are at selecting the best proposals to fund. Even experts can make mistakes, allocate insufficient time to the task, have unreasonable biases or partialities, have little knowledge of some areas of the application, or be inexpert on the application overall. Nevertheless, as discussed in the introduction and given that there is no real choice, it seems reasonable to at least believe that field experts are better than LLMs or at least should be prioritised over LLMs since they have (or should have) some genuine specialist academic knowledge rather than being clever pattern matchers. To illustrate the difference, human experts might be more open to funding proposals addressing an emerging challenge (e.g., a new virus, wide uptake of GenAI) and considering broader issues (e.g., the need to protect diversity or promising young researchers within a field), so the long term effect of entirely replacing humans with LLMs might be to undermine the ability of academic research to respond to challenges and an overall weaker health due to patchy growth. Of course, without AI input, fields can also encourage weak or irrelevant research (e.g., Koskela 2017; Tourish 2020), so this potential is not unique to AI. A more fundamental limitation is that commercial LLMs can be influenced by the political priorities of their creators or in the data that they are trained on, potentially allowing non-academics to influence decisions about funding research into sensitive topics, such as gender identity (e.g., Ovalle et al. 2023; Voutyrakou and Voutyrakos 2026). These issues, and the importance of funding, suggest that humans need to take responsibility for important decisions (e.g., Jovchevski et al. 2026).

For RQ1, the correlation between the proposal rankings produced by LLM scores and expert scores varied substantially between models, calls and prompt configurations. The strongest overall configuration, Gemma 3 27B with the long and short prompts combined, had a moderate mean correlation with reviewer Scores (rho = 0.263) and a weak mean correlation with panel-member Prescores (rho = 0.165). Thus, this configuration showed some ability to reproduce the proposal rankings implied by expert assessments, with stronger alignment to initial reviewer Scores than to panel member Prescores (on a high scoring subset). The smaller Gemma 3 variants and the two reasoning models generally performed less well. The Gemma 3 27b correlations in the current case (0.26 from 10 iterations, 0.24 from five) are similar to the correlation of 0.24 found with Gemma 3 27b (five iterations averaged) for full text Swedish medical proposals from 1994 (Sandström and Thelwall 2026). Given that a stronger correlation was found for title/summary input for the Swedish case, this suggests that higher correlations could also be obtained for UKRI proposal summaries, although these were not available. The correlation is towards the lower end of the correlations for full text physics equipment applications previously investigated, which ranged between 0.2 and 0.8 for different years and tools (Figure 4 of: Ding et al. 2025), but this is a specialised type of application.

The relatively low correlations between LLM scores and mean reviewer scores are nevertheless higher than the correlations between individual reviewers for the four calls, reflecting the general pattern that there is often little correlation between individual reviewer scores for grant applications (e.g., Fogelholm et al. 2012; Hodgson 1995; Pier et al. 2018). Nevertheless, if the best performing LLM is compared against the individual reviewer scores (rather than the mean of the reviewer scores) then the reviewers correlate with each other individually more strongly than the LLM correlates with the individual reviewers (Table 2), with an average correlation of 0.186. Thus, LLMs do not seem to be quite as strong as individual reviewers from a technical accuracy (correlation) standpoint, even though the differences are not large. The difference is even sharper for panellist prescores (Table 3), where the mean correlation between Gemma 3 27B and individual panellists is 0.144.

Table 2. Spearman correlations between scores from Gemma 3 27B (both prompts combined) and reviewer scores from the first three review sets and the mean of these three corelations. The final column is the mean of the Spearman correlations between scores from the three review sets (the grey bars in Figure 1).

| Call | Review set 1 | Review set 2 | Review set 3 | LLM mean | | (Reviewer vs. reviewer) mean |
|---|---|---|---|---|---|---|
| EPSRC 1 | 0.204 | 0.212 | 0.172 | 0.196 | | 0.242 |
| ESRC 1 | 0.143 | 0.193 | 0.174 | 0.170 | | 0.192 |
| ESRC 2 | 0.183 | 0.194 | 0.162 | 0.179 | | 0.232 |
| EPSRC 2 | 0.180 | 0.182 | 0.231 | 0.198 | | 0.301 |

Table 3. Spearman correlations between scores from Gemma 3 27B (both prompts combined) and panellist prescores from the first three review sets and the mean of these three corelations. The final column is the mean of the Spearman correlations between prescores from the three review sets (the blue bars in Figure 1).

| Call | Review set 1 | Review set 2 | Review set 3 | LLM mean | | (Panellist vs. panellist) mean |
|---|---|---|---|---|---|---|
| EPSRC 1 | 0.233 | 0.110 | 0.198 | 0.181 | | 0.512 |
| ESRC 1 | 0.003 | 0.020 | 0.206 | 0.076 | | 0.225 |
| ESRC 2 | 0.177 | 0.078 | 0.115 | 0.124 | | 0.301 |
| EPSRC 2 | 0.191 | 0.201 | 0.187 | 0.193 | | 0.488 |

Although the Spearman correlations do not have a simple intuitive interpretation, Kendall's tau-b correlations were also calculated for this purpose. For Gemma 3 27B with long and short prompts combined, Kendall's tau-b correlations varied between 0.15 and 0.22. This means that, if the human mean scores are correct, a decision maker picking the best proposal from a pair on the basis of Gemma 3 27B scores would be "correct" (in the sense of giving the same result as the selected human experts) between (1+0.15)/2 = 57% and (1+0.22)/2 = 61% of the time. This is a relatively modest improvement on guessing (i.e., 50%), but perhaps comparable to existing procedures (e.g., panel discussions), given the high rate of expert disagreement. Moreover, the real improvement over guessing may be higher given that the mean reviewer scores may also be incorrect.

For RQ2, the correlations varied across the four funding calls. For the strongest configuration, Gemma 3 27B with the long and short prompts combined, the observed Spearman correlations with both reviewer Scores and panel-member Prescores were higher in the two EPSRC calls than in the two ESRC calls. This pattern may partly reflect the higher agreement among experts in the EPSRC calls, which provides a more consistent benchmark against which to compare LLM scores. The greater dispersion of EPSRC Prescores may also have made correlations with Prescores easier to detect. However, the patterns for the other models and prompt configurations were less consistent. The results therefore indicate variation across calls, but further research comparing more calls within and across a wider range of research fields is needed to establish whether, and how, research field affects the alignment between LLM and expert scores, and to distinguish field effects from call-specific differences.

# 5 Conclusions

The low but positive and statistically significant rank correlations between LLM average scores and reviewer scores suggest that LLMs may usefully play a role in contemporary funding decisions. Whilst they may not be accurate enough to play an important role, and as argued above replacing human experts with AI seems undesirable (Sikimić 2025), they might still be used in situations where higher accuracy is not available, or when efficiency can be improved without compromising the overall outcome. These include the triage role pioneered by La Caixa (Carbonell Cortés et al. 2024), replacing one reviewer in the initial review stage, as a second opinion for bias checks (although LLMs also have biases: Thelwall and Kurt 2025), and identifying outliers where all reviewers might be wrong so the proposal might need to be revisited. Given the comparable rate of agreement with human reviewers, there may even be exceptional cases where AI decisions might be accepted on their own, such as for emergency funding, for small funding proposals where the expert time needed to assess them would be too costly relative to the value of the grants, or possibly even as an alternative to random selection in cases where this is currently used.

If LLMs are used to support grant funding decisions then it is important to recognise the weakness of the evidence that they give, allowing human expert opinions to be the primary source unless there is a strong reason not to. In addition, systemic effects should be considered, for example applicants writing their proposals for AI rather than experts, and potential long-term damage to a field if research is too strongly influenced by LLM decisions.

From a funder perspective, they are ultimately responsible for the decision-making process (i.e., the "humans" in: Jovchevski et al. 2026), with their stakeholders including the government (who supply the money) and academics (who spend the money). Under the current UK system, although UKRI has designed the funding approach (i.e., independent reviewers, then a funding panel; the nature of the funding calls) and manage it (e.g., selecting reviewers), they largely delegate the outcome to academics in the form of the selected reviewers and panel members. Increasing the role of LLMs would shift the decision responsibility partly (if LLMs support reviewers) or fully (if LLMs replace reviewers) from academics to the funding councils. In practice, this suggests that the funding councils would need to exercise this responsibility by taking steps to protect against unintended consequences, such by monitoring for LLM biases and unintended consequences. Funders may also need to reassure their stakeholders: although government may welcome LLM-based efficiency gains, they will need to be reassured that their money is still being allocated responsibly; and academics might welcome less reviewing and/or faster decisions but may need evidence that their fields are not undermined by the new processes. Given the reluctance of UK academics to allow AI to evaluate them (Watermeyer et al. 2025), and a recent academic funding-related public controversy (Johnson 2026), a cautious and reflexive approach to LLM adoption seems desirable.

# 6 Acknowledgement

This project was funded by the Economic and Social Research Council (ESRC), UK (UKRI2101): LLMs Supporting Grant Peer Review. Dr Ben Steyn, Co-Head of the UK Metascience Unit, and Dr Cillian Brophy, Senior Analyst in the UK Metascience Unit provided comments on an earlier version. The opinions expressed in this article do not necessarily reflect the reviews of the funder.

# 8 Appendix

## *8.1 EPSRC long system prompt*

You are an expert academic reviewer that carefully assesses applications for research funding according to the following guidelines.
Vision and Approach
Assess how the proposed work:
* is of excellent quality and importance within or beyond the field(s) or area(s)
* has the potential to advance current understanding, generates new knowledge, thinking or discovery within or beyond the field or area
* is timely given current trends, context and needs
* impacts world-leading research, society, the economy or the environment

Within the Vision section assess how the proposed work:
* identifies the potential direct or indirect benefits and who the beneficiaries might be

For the approach, assess the extent to which the proposal:
* is effective and appropriate to achieve the objectives
* is feasible, and comprehensively identifies any risks to delivery and how they will be managed
* if applicable, uses a clear and transparent methodology
* if applicable, summarises the previous work and describes how this will be built upon and progressed

* will maximise translation of outputs into outcomes and impacts
* describes how the applicant's, and if applicable their team's, research environment (in terms of the place, and relevance to the project) will contribute to the success of the work
Also assess the extent to which the proposal
* demonstrates access to the appropriate services, facilities, infrastructure, or equipment to deliver the proposed work.

Applicant and team capability to deliver
Assess the extent to which the applicant, and if relevant their team, have:
* the relevant experience (appropriate to career stage) to deliver the proposed work
* the right balance of skills and expertise to cover the proposed work
* the appropriate leadership and management skills to deliver the work and your approach to develop others
* contributed to developing a positive research environment and wider community

Resources and cost justification
Assess the extent to which the resources for the proposed work:
* are comprehensive, appropriate, and justified
* represent the optimal use of resources to achieve the intended outcomes
* maximise potential outcomes and impacts
Do not look for detailed costs or a line-by-line breakdown of all project resources. Do not consider the level of matched university funding as a factor.

Ethics and responsible research and innovation
Assess the extent to which the proposal covers:
* the relevant ethical or responsible research and innovation considerations
* how they will be managed

### *8.2 EPSRC short system prompt*

You are an expert academic reviewer that carefully evaluates applications for research funding. Assess the following areas related to the application questions:
* vision of the project
* approach to the project
* capability of the applicant or applicants and the project team to deliver the project
* resources requested to do the project
* ethical and responsible research and innovation considerations of the project

### *8.3 ESRC long system prompt*

You are an expert academic reviewer that carefully assesses applications for research funding according to the following guidelines.

Vision
Assess how the proposed work:
* is of excellent quality and importance within or beyond the fields or areas
* has the potential to advance current understanding, or generate new knowledge, thinking or discovery within or beyond the field or area

* is timely given current trends, context, and needs
* impacts world-leading research, society, the economy, or the environment

Also assess:
* potential beneficiaries and users of the proposed research, including the relevance of the research to these beneficiaries
* the expected outputs; both academic and those orientated to users

Approach
Assess how far the proposed work:
* is effective and appropriate to achieve the objectives
* is feasible, and comprehensively identifies any risks to delivery and how they will be managed
* uses a clearly written and transparent methodology (if applicable)
* summarises the previous work and describes how this will be built upon and progressed (if applicable)
* will maximise translation of outputs into outcomes and impacts
* describes how the applicant's, and if applicable the team's, research environment (in terms of the place and relevance to the project) will contribute to the success of the work

Also assess:
* both the framework and specific analysis methods proposed and the reasons for their choice. Value particularly any innovation in this or how different methodologies or methods may be combined.
* the steps to provide opportunities for users to benefit from the research, and to ensure that the research has maximum economic and societal impact

Applicant and team capability to deliver
Assess how far the applicant, and if relevant their team, have:
* the relevant experience (appropriate to career stage) to deliver the proposed work
* the right balance of skills and expertise to cover the proposed work
* the appropriate leadership and management skills to deliver the work and approach to develop others
* contributed to developing a positive research environment and wider community

Ethics and responsible research and innovation (RRI)
Assess the extent to which the application identifies and evaluates:
* the relevant ethical or responsible research and innovation considerations
* how these considerations will be managed

Resources and cost justification
Assess the extent to which the resources for the proposed work:
* are comprehensive, appropriate, and justified
* represent the optimal use of resources to achieve the intended outcomes
* maximise potential outcomes and impacts
Do not look for detailed costs or a line-by-line breakdown of all project resources. Do not consider the level of matched university funding as a factor.

## 8.4 ESRC short system prompt

You are an expert academic reviewer that carefully evaluates applications for research funding. Assess the following areas related to the application questions:
* vision
* approach
* data management and sharing
* applicant and team capability to deliver
* ethics and responsible research and innovation (RRI)
* resources and cost justification

## 8.5 User prompt common scoring segment

* 6 Exceptional: the application is outstanding. It addresses all of the assessment criteria and meets them to an exceptional level.
* 5 Excellent: the application is very high quality. It addresses most of the assessment criteria and meets them to an excellent level. There are very minor weaknesses.
* 4 Very good: the application demonstrates considerable quality. It meets most of the assessment criteria to a high level. There are minor weaknesses.
* 3 Good: the application is of good quality. It meets most of the assessment criteria to an acceptable level, but not across all aspects of the proposed activities. There are weaknesses.
* 2 Weak: the application is not sufficiently competitive. It meets some of the assessment criteria to an adequate level. There are, however, significant weaknesses.
* 1 Poor: the application is flawed or of unsuitable quality for funding. It does not meet the assessment criteria to an adequate level.

## 8.6 The five user prompt varying segments

Assess the application below and give it an overall score between 1 and 6 according to the following criteria, using fractions if necessary. Take into account that it was written in 2023.
[common scoring segment]
Your report should be strictly in the following format.
Written assessment:
Score:
###
[proposal text]

Evaluate the application below and give an overall score between 1 and 6 with the following criteria, using fractions when necessary. Make allowance for the fact that it was written in 2023.
[common scoring segment]
The report should strictly follow this format.
Written assessment:
Score:
###
[proposal text]

Judge the funding grant application below and allocate an overall score between 1 and 6 according to the following criteria, using fractions if and when necessary. Take into account that it was submitted in 2023.
[common scoring segment]
Write the report strictly in the following format.
Written assessment:
Score:
###
[proposal text]

Analyse the funding application below and decide on an overall score from 1 to 6 according to these criteria, with fractions if necessary. Take into account that the document was written in 2023.
[common scoring segment]
Keep the report strictly in the following format.
Written assessment:
Score:
###
[proposal text]

Critically evaluate the funding application below and assign it an overall score using the range 1 to 6 according to the following criteria. Use fractions if necessary. Take into account that the proposal is from 2023.
[common scoring segment]
The report should adhere strictly to this format.
Written assessment:
Score:
###
[proposal text]